\def\a{\alpha}\def\d{\delta}\def\e{\epsilon}
\def\f{\phi}\def\g{\gamma}
\def\l{\lambda}\def\m{\mu}\def\n{\nu}\def
\p{\pi}\def\q{\psi}\def\r{\rho}\def\t{\tau}
\def\v{\varphi}
\def\ee{\varepsilon}

\def\de{\partial}
\def\mo{{-1}}\def\ha{{1\over 2}}

\def\({\left(}\def\){\right)}\def\[{\left[}\def\]{\right]}

\def\coo{coordinates }

\def \schr{Schr\"odinger }
\def\KG{Klein-Gordon }

\def\wrt{with respect to }\def\ie{i.e.\ }
\def\eom{equations of motion }

\def\section#1{\bigskip\noindent{\bf#1}\smallskip}
\def\subsect#1{\bigskip\noindent{\it#1}\smallskip}
\def\nota{\footnote{$^\dagger$}}

\def\PR#1{Phys.\ Rev.\ {\bf#1}}

\def\ref#1{\medskip\everypar={\hangindent 2\parindent}#1}
\def\beginref{\begingroup
\bigskip
\centerline{\bf References}
\nobreak\noindent}
\def\endref{\par\endgroup}

\def\cL{{\cal L}}\def\cH{{\cal H}}
\def\xds{\sqrt{(\dot x^\r)^2}}\def\xdt{\sqrt{1-x_k'^2}}
\def\repa{reparametrization }

%\magnification=1200

{\nopagenumbers
\line{}
\vskip20pt
\centerline{\bf Analytical mechanics of a relativistic particle in a positional potential}

\vskip20pt
\centerline{{\bf S. Mignemi}\nota{e-mail: smignemi@unica.it}}
\vskip10pt
\centerline {Dipartimento di Matematica, Universit\`a di Cagliari}
\centerline{viale Merello 92, 09123 Cagliari, Italy}
\smallskip
\centerline{and INFN, Sezione di Cagliari}
\vskip100pt
\centerline{\bf Abstract}
\vskip10pt
{\noindent We propose a form for the action of a relativistic particle subject to a
positional force that is invariant under time reparametrization and therefore allows for
a consistent Hamiltonian formulation of the dynamics.
This approach can be useful in the study of phenomenological
models. Also the Dirac and \KG equation differ from the standard formulation, with corrections
of order $(E-m)/m$ in the energy spectra.
}
\vskip280pt
%P.A.C.S. Numbers:
\vfil\eject}

\section{1. Introduction}
Analytical mechanics is at the basis of modern theoretical physics. In
particular, Hamiltonian methods are largely employed in several applications
and are essential for the formulation of the quantum theory.
However, not all systems have been consistently described in Hamiltonian
terms.
For example, if one excludes fundamental interaction, like Maxwell theory,
whose coupling to particles is dictated by gauge invariance,
in special relativity it is not easy to
introduce particle interactions that preserve the Lorentz and the
\repa invariance of the action.
While for phenomenological interactions the request of Lorentz invariance is
not compelling, the breakdown of the \repa invariance has serious
consequences on the possibility of defining a consistent Hamiltonian
for the model under study. The lack of a Hamiltonian formulation
also implies that in quantum theory the \KG or Dirac equations for generic
interactions have to be
postulated rather than derived from the correspondence principle.

These problems arise especially for models like that of an external central
force acting on a particle, when the force is fixed a priori and is not
determined by field equations derived from a variational principle.
An important case is that of the harmonic oscillator, whose relativistic
formulation is problematic [1].
Usually, in such cases the Hamiltonian is defined only for a specific choice
of the time coordinate.

In this paper, we propose a different solution that, slightly modifying the
coupling, makes the action \repa invariant. This allows to obtain a Hamiltonian
by means of the usual Dirac formalism for constrained systems.
The corrections with respect to the standard formalism are of order $(E-m)/m$,
where $E$ is the relativistic energy and $m$ the mass of the particle.
Moreover, a really covariant \KG or Dirac equation can be obtained.
In the case of the harmonic oscillator, a proposal leading to equivalent results
was advanced long ago in [1], starting from different considerations.

The paper is organized as follows: in sect.\ 2 we review the Lagrangian and
Hamiltonian formalism for a free particle. In sect.\ 3 we describe our proposal
and compare it with the standard formulation of the relativistic interacting
particle.
In sect.\ 4 we give some elementary instances of application of the formalism.
In sect.\ 5 the changes in the \KG equations are discussed and illustrated with
simple examples in sect.\ 6.

\section{2. Free particle}
In special relativity, the motion of a free particle in \coo $x^\m$ is determined
by the variation of the action
$$S=\int\cL\,d\t=-m\int\xds\,d\t=-m\int ds,\eqno(1)$$
where $\cL$ is the Lagrangian density, $\t$ an arbitrary evolution parameter,
a dot denotes a derivative \wrt $\t$, $m$ is the rest mass and $s$ the proper
time on the trajectory.
The action is invariant under Lorentz transformations and under
reparametrizations $\t\to\t'(\t)$. The Euler-Lagrange equations for
the action (1) read
$$m\,{d\over d\t}\,{\dot x^\m\over\xds}=0.\eqno(2)$$

The dynamics can also be written in Hamiltonian form [2].
The momentum conjugated to $x^\m$ is defined as
$$p_\m={\de\cL\over\de\dot x^\m}=m{\dot x^\m\over\xds}=m{dx^\m\over ds},\eqno(3)$$
and satisfies the constraint
$$p_\m^2=m^2.\eqno(4)$$

Since the action is \repa invariant, the Hamiltonian $\cH$ vanishes, and the
action can be written in first-order form as
$$S=\int d\t(\dot x^\m p_\m-\l\cH),\qquad\cH=p_\m^2-m^2,\eqno(5)$$
where $\l$ is a Lagrange multiplier enforcing the constraint (4).

The Hamilton equations ensuing from the action (5) are
$$\dot x_\m=\{x_\m,\l\cH\}=2\l p_\m,\qquad\dot p_\m =\{p_\m,\l\cH\}=0,\eqno(6)$$
from which, comparing with (3), follows that
$$\l={\xds\over2m}.\eqno(7)$$

\medskip
In Dirac formalism, the constraint (4) is first order. One can therefore
reduce the system by choosing a gauge, \ie fixing the time coordinate.

The standard choice is $t=x^0$, namely one identifies the evolution time $t$
with the coordinate time $x_0$. Since $\{t,\l\cH\}=2\l p_0$, no secondary
constraints arise and one easily checks that
$$x_i'={p_i\over p_0}={p_i\over\sqrt{p_k^2+m^2}},\qquad p_i'=0,
\qquad\l={1\over2m}\xdt={1\over2p_0}\eqno(8)$$
where $i=1,2,3$ and a prime denotes a derivative \wrt $t$. The reduced action is
then
$$S=\int dt\(x'_ip_i-\sqrt{p_i^2+m^2}\),\eqno(9)$$
and the 3-dimensional effective Hamiltonian $H=p_0=\sqrt{p_i^2+m^2}$ can be identified
with the energy of the particle in the laboratory frame.

Other choices of the gauge are however possible. For example, with the choice
$t=x^\m p_\m$, the evolution time coincides with the proper time of the particle.

\section{3. Particle in an external potential}
The addition of an external potential acting on the free particle presents some
problems and to our knowledge has not been discussed in depth.
In the standard formalism, one simply adds to the action a potential term
$$S_{int}=-\int V(x,\dot x)\,d\t.\eqno(10)$$
In general the potential breaks the Lorentz invariance. Moreover, unless $V$
is homogeneous of degree one in the velocity, as in the Maxwell case when
$V=\dot x^\m A_\m(x)$,
$S_{int}$ is not invariant under reparametrization, and the Dirac formalism cannot be
consistently applied for writing down the \eom in Hamiltonian form.

In the following we discuss the case of positional potentials $V=V(x)$ that do
not depend on the velocity of the particle. These usually arise as phenomenological
potentials, like those associated to elastic forces. For positional potentials,
the interaction term (10) leads to the standard \eom
$${d\over d\t}\ {m\,\dot x^\m\over\xds}- \de^\m V=0.\eqno(11)$$
Multiplying eq.\ (11) by $\dot x_\m$ and integrating, one can show that the total
energy
$$E={m\over\xds}+V\eqno(12)$$
is conserved. Since $V=V(x)$, the momenta are given by (3) as for the free particle
and satisfy the constraint (4).
If one tries to apply the Dirac formalism to the action so defined, one obtains that
$V$ must be a constant.
This is a consequence of the lack of reparametrization invariance of (10).

\medskip
In order to avoid these problems and to preserve \repa invariance, we propose
an interaction term of the form
$$S_{int}=-\int\xds\ V(x)\,d\t,\eqno(13)$$
leading to the total action
$$S=-\int\xds\,(m+V)\,d\t.\eqno(14)$$
The Euler-Lagrange equations for the action (14) read then
$${d\over d\t}\ {(m+V)\,\dot x^\m\over\xds}-\xds\ \de^\m V=0.\eqno(15)$$
If $V$ does not depend on $\dot x^\m$, eq.\ (15) can also be written
$$(m+V){d\over d\t}\ {\dot x^\m\over\xds}+{\dot x^\m\dot x^\n\de_\nu V-
(\dot x^\n)^2\de^\m V\over\xds}=0.\eqno(16)$$
One can again find the energy integral by multiplying eq.\ (15) by $\dot x_\m$ and
integrating, obtaining
$$E={m+V\over\xds}.\eqno(17)$$

To pass to the Hamiltonian formulation, we compute the momentum conjugate to $x^\m$,
$$p_\m={\de\cL\over\de\dot x^\m}=(m+V){\dot x^\m\over\xds}=(m+V){dx^\m\over ds},
\eqno(18)$$
which is subject to the constraint
$$p_\m^2=(m+V)^2.\eqno(19)$$
It follows that the action can be written in Hamiltonian form as
$$S=\int d\t(\dot x^\m p_\m-\l\cH),\qquad\cH=p_\m^2-(m+V)^2\eqno(20)$$
where $\l$ is a Lagrange multiplier enforcing the constraint (19).

The Hamilton equations read
$$\dot x_\m=\{x_\m,\l\cH\}=2\l p_\m,\qquad\dot p_\m =\{p_\m,\l\cH\}=-2\l(m+V)\de_\m V,
\eqno(21)$$
from which follows that
$$\l={\xds\over2(m+V)}.\eqno(22)$$

In our formalism, the potential essentially plays the role of a variable mass added to
the rest mass of the particle. This is in accord with special relativity, since the
external potential is fixed and does not change with the motion of the particle.
So its field cannot contribute to the energy balance in any other way than effectively
modifying the mass of the particle.

\medskip
Again, one can choose a gauge in order to reduce to a three-dimensional problem
with external time. Taking $t=x^0$, one has $\{t-x_0,\l\cH\}=2\l p_0$, and hence no
secondary constraints arise if the potential does not depend on $x_0$. Moreover
$$x_i'={p_i\over p_0}={p_i\over\sqrt{p_i^2+(m+V)^2}},\qquad p_i'=-{m+V\over p_0}\ \de_i V,
\qquad\l={\xdt\over2(m+V)}={1\over2p_0}.\eqno(23)$$
In this case the reduced action is
$$S=\int dt\(x'_ip_i-\sqrt{p_i^2+(m+V)^2}\).\eqno(24)$$
The 3-dimensional Hamiltonian $H=\sqrt{p_i^2+(m+V)^2}$  represents
the energy in the laboratory frame. It should be compared with the expression adopted
in the standard formalism, namely $H=\sqrt{p_i^2+m^2}+V$.
This is obtained from the reduced lagrangian of the standard formalism in the gauge $t=x_0$
and has no covariant meaning. The two Hamiltonians differ by terms of order $(E-m)/m$.

For $m\gg p_i$, $m\gg V$, the expansion of both Hamiltonians gives rise to the classical
expression plus higher-order corrections,
$$E\sim m+{p_i^2\over2m}+V+\dots\eqno(25)$$
However, with our Hamiltonian higher-order corrections are present in the expansion also
for the potential term.

Like for the free particle, also in presence of a potential term, one can choose a different
gauge. For example, in 1+1 dimensions, the evolution time can be made to coincide with the
proper time of the particle for the choice $t=(m+V)\(1+\int^x{mdx\over(m+V)^3}\)p$.

\section{4. Examples}
We give two elementary examples that we solve in the Lagrangian formalism,
and compare them with the results obtained using the standard formalism.
The potentials are not Lorentz invariant and therefore the results
hold only in the rest reference frame. However the \eom are \repa invariant.

\subsect{4.1 Harmonic oscillator}
For a 1-dimensional system, the \eom are immediately integrated.
Choosing $t=x_0$, the conserved energy (17) can be written
$$E={m+V\over\sqrt{1-{x'}^2}}.\eqno(26)$$
Easy calculations lead to the differential equation
$$x'={1\over E}\ \sqrt{E^2-(m+V)^2}.\eqno(27)$$
that can be immediately integrated.
In the case of a harmonic oscillator, $V=\ha kx^2$, the same equation has been
obtained in [1], starting from a different perspective.

The solution of (27) is given by [1]
$$x=\sqrt{E^2-m^2\over kE}\ {\rm sd}\!\!\(\sqrt{k\over E}\ t,{E-m\over2E}\),\eqno(28)$$
with period
$$T=4\sqrt{E\over k}\ {\rm K}\!\!\({E-m\over2E}\)\sim2\p\sqrt{m\over k}
\(1+{5\over8}\ {E-m\over m}+\dots\),\eqno(29)$$
where sd and K are elliptic functions.
Contrary to nonrelativistic mechanics, the period of the oscillations is not constant,
but depends on the energy.
We have written down the first term of its expansion in $(E-m)/m$ around the
nonrelativistic value.

Note that in the standard formalism, the equation of motion would have read
$$x'={\sqrt{(E-V)^2-m^2}\over E-V}\eqno(30)$$
Also this equation can be solved in terms of elliptic functions [3].
In particular, the period of the solution is
$$T\sim2\p\sqrt{m\over k}
\(1+{3\over8}\ {E-m\over m}+\dots\)\eqno(31)$$
whose dependence on energy differs from (29).

\subsect{4.2 Kepler problem}
It is well known that the Kepler problem can be reduced to a one-dimensional problem
because of the presence of two integrals of the motion related to the conservation of
the angular momentum. This implies that the motion occurs on a plane and that the
component of the angular momentum orthogonal to the plane is conserved.
Taking $t=x_0$ and polar \coo $r$ and $\v$ on the plane of the motion, from (17) one has
$$E={m+V\over\sqrt{1-{r'}^2-r^2{\v'}^2}},\eqno(32)$$
while from the conservation of the norm of the angular momentum follows that
$$l={mr^2\v'\over\sqrt{1-{r'}^2-r^2{\v'}^2}}\eqno(33)$$
is a constant. After standard computations from (32) and (33) one obtains the equation
of the orbits,
$$\({dr\over d\v}\)^2+r^2={r^4\over l^2}\ [E^2-(m+V)^2]\eqno(34)$$
Defining $u={1\over r}$ and $V=-{\a\over r}=-\a u$, eq.\ (34) becomes
$$\({du\over d\v}\)^2+u^2={1\over l^2}\ [E^2-(m-\a u)^2]\eqno(35)$$
whose solution is given by
$$r={1\over u}={p\over1+\ee\cos q(\v-\v_0)},\eqno(36)$$
with
$$q=\sqrt{1+{\a^2\over l^2}},\qquad p={l^2+\a^2\over\a m},\qquad
\ee={E\over m}\ \sqrt{1+{(E^2-m^2)\,l^2\over\a^2E^2}}.\eqno(37)$$

These values should be compared with those obtained using the standard formalism,
namely
$$q=\sqrt{1-{\a^2\over l^2}},\qquad p={l^2-\a^2\over\a E},\qquad
\ee={m\over E}\ \sqrt{1+{(E^2-m^2)\,l^2\over\a^2m^2}}.\eqno(38)$$
In the formulae above, $\ee$ is the eccentricity of the orbit, while
the parameter $p$ is related to the size of the orbit and $q$ to the perihelion advance
$\d=2\p(q^\mo-1)$.
The angle $\d$ has opposite value in our formalism \wrt the standard one. Of course,
the correct value is that given by general relativity and is three times the standard one.
The other parameters of the orbit calculated in our formalism differ as usual for terms
of order $(E-m)/m$ from the standard ones.

\section{5. Quantization}
A first quantized relativistic equation can be obtained as usual
starting from the Hamiltonian (20). The \KG equation is obtained with the
substitution $p_\m\to i\hbar\de_\m$, namely
$$\hbar^2\({\de^2\f\over\de t^2}-{\de^2\q\over\de x_i^2}\)+(m+V)^2\q=0.\eqno(39)$$
Setting $\q=\f\,e^{-iEt/\hbar}$, it reduces to
$$-\hbar^2{\de^2\f\over\de x_i^2}+[(m+V)^2-E^2]\f=0,\eqno(40)$$
that has the form of a \schr equation with $E_{eff}=E^2-m^2$, $V_{eff}=2mV+V^2$.
Note that if the potential $V$ is positive definite, also $V_{eff}$ is. This is not
true in the standard formalism, leading to problems with stability.

In fact, in the standard formalism, the \KG equation is assumed to be
$$-\hbar^2{\de^2\q\over\de x_i^2}+[m^2-(E-V)^2]\f=0.\eqno(41)$$
It also has the form of a \schr equation with $E_{eff}=E^2-m^2$, $V_{eff}=2EV-V^2$.
Note that this equation cannot be derived from a classical Hamiltonian, since that
is not well defined. Moreover, the presence of $E$ in the effective potential makes
it difficult to find the spectrum of the energy explicitly.

Finally, we notice that also the Dirac equation can be modified in accord with (39),
as
$$[i\hbar\g^\m\de_\m-(m+V)]\q=0.\eqno(42)$$

\section{6. Examples}
We give two simple examples of solution of the \KG equation. In the first case, only
a perturbative solution is possible, while the second is constructed so that an exact
solution can be found.
For simplicity, in this section we set $\hbar=1$.
\subsect{6.1. Quantum harmonic oscillator}
The \KG equation (39) for a one-dimensional harmonic oscillator with potential
$V=\ha kx^2$ reads
$${d^2\f\over dx^2}=(\m x^4+mk\,x^2-\e)\f,\eqno(43)$$
where $\e=E^2-m^2$, $\m=k^2/4$, and has the form of a nonrelativistic
biquadratic oscillator. Its energy spectrum can be obtained using standard
perturbation theory, taking the quartic term as a perturbation of the
nonrelativistic harmonic oscillator. The calculation has been performed in [1]
and gives at first order
$$E_n=\pm\[m+\sqrt{k\over m}\(n+\ha+{3\over4}\,\m\(n^2+n+\ha\)\)\],\eqno(44)$$
where $n$ is an integer.

In the standard formalism, the \KG equation (38) reads instead
$${d^2\f\over dx^2}=(-\m x^4+Ek\,x^2-\e)\f,\eqno(45)$$
with $\e$ and $\m$ as before,
and the effective potential is no longer positive definite, leading to possible
instabilities [4]. Neglecting this problem, one can obtain as before the energy
spectrum, which at first order in $\hbar$, and neglecting corrections of order
$(E-m)/m$, yields the opposite sign for the correction, namely
$$E_n=\pm\[m+\sqrt{k\over m}\(n+\ha-{3\over4}\,\m\(n^2+n+\ha\)\)\].\eqno(46)$$
\vfil\eject
\subsect{6.2. Exactly solvable potential}
An instance in which a solution of the \KG equation (39) can be found explicitly is
$$V=m\({1\over\cos x}-1\).\eqno(47)$$
This is a potential well whose value is zero at the origin and infinite
at $x=\pm{\p\over2}$. The \KG equation reads
$${d^2\f\over dx^2}-\[{m^2\over\cos^2x}-E^2\]\f=0.\eqno(48)$$
Defining $z=\sin x$, eq.\ (48) can be put in the form
$${d^2\f\over dz^2}-z\,{d\f\over dz}-\[{m^2\over(1-z^2)^2}-{E^2\over1-z^2}\]\f=0,
\eqno(49)$$
that in turn can be reduced to a standard hypergeometric equation with solution
$$\f=\cos^\n\!x\ {\rm F}\!\!\(\n+E,\n-E,\n+\ha,{1+\sin x\over2}\),\eqno(50)$$
where F is a hypergeometric function and $\n=\ha\(1+\sqrt{1+4m^2}\)$.
Imposing the vanishing of $\f$ at $x=\pm{\p\over2}$, one finds the eigenfunctions
$$\f_n=\cos^\n\!x\ C_n^{(\n)}(\sin x),\eqno(51)$$
with $C_n^{(\n)}$ Chebyshev polynomials, and the energy spectrum
$$E=\pm\[m^2+\(n+\ha\)\(1+\sqrt{1+4m^2}\)+n^2\]^{1/2},\eqno(52)$$
where $n$ is an integer.

In the standard formalism, the \KG equation cannot be solved analytically and
therefore we do not pursue its investigation here.
\section{7. Conclusions}
We have shown that it is possible to formulate the problem of the motion of a
particle in a positional potential in special relativity in such a way to preserve
\repa invariance, and hence to give a consistent Hamiltonian formulation in terms
of the Dirac formalism for constrained systems.
Of course, potentials of the kind investigated in this paper do not occur in
fundamental interactions, whose action is already invariant under reparametrization,
but can be of interest for phenomenological models.

\beginref
\ref [1] A.L. Harvey, \PR{D6}, 1474 (1972).
\ref [2] A. Hanson, T. Regge and C. Teitelboim, {\sl "Constrained Hamiltonian systems"},
Accademia Nazionale dei Lincei, Rome 1976.
\ref [3] L.A. MacColl,  Am. J. Phys. {\bf 25}, 535 (1957).
\ref [4] P.O. Lipas, Am. J. Phys. {\bf 38}, 85 (1970).

\endref
\end
Another possibility is to choose $t={x^\m p_\m\over m}$.
In this case one easily checks that
$$x_i'={p_i\over m},\qquad p_i'=0,\qquad\l={1\over2m}.\eqno(10)$$
Comparing with (), it is evident that in this gauge $t$ coincides with
the proper time $s$ along the trajectory. In this case the reduced action is
$$S=\int dt\[\dot x_ip_i-{1\over m}\(p_i^2+m^2\)\],\eqno(11)$$
and $H={1\over m}\(p_i^2+m^2\)$ assumes the nonrelativistic form.

A second possibility is to choose $t={x^\m p_\m\over m}$.
In this case one easily checks that
$$x_i'={mp_i\over(m+V)(m+V-x^\m\de_\m V)},\qquad p_i'=-{\de_i V\over m+V-x^\m\de_\m V},
\qquad\l={1\over2(m+V-x^\m\de_\m V)},\eqno(23)$$
and  the reduced action reads
$$S=\int dt\[x'_ip_i-m{p_i^2+(m+V)^2\over(m+V)(m+V-x^\m\de_\m V)}\].\eqno(24)$$
In this case the reduced Hamiltonian differs from its classical equivalent.

Usually these interactions break the Lorentz invariance (for example they
depend on the spatial distance or have a fixed center). What is more serious
from the point of view of analytical mechanics is that the usual formulation
obtained by simply adding to the action the potential energy,
is not reparametrization invariant, and hence a consistent Hamiltonian
formulation is not possible. In this paper we propose a simple modification
of the usual formalism that solves this problem.